\documentclass[12pt,epsf]{article}
\usepackage{hyperref}
\usepackage{graphics}
\usepackage{amssymb}
\usepackage{epsf}
\usepackage{epsfig}

\def\ds#1{#1\kern-1ex\hbox{/}}
\def\dsh{h\kern-1.2ex /}

\newcommand{\bea}{\begin{eqnarray}}
\newcommand{\eea}{\end{eqnarray}}

\def\beq{\begin{equation}}
\def\eeq{\end{equation}}

\def\beqn{\begin{eqnarray}}
\def\eeqn{\end{eqnarray}}
\def\ba{\begin{eqnarray}}
\def\ea{\end{eqnarray}}

\newcommand{\beqa}{\begin{eqnarray}}
\newcommand{\eeqa}{\end{eqnarray}}

\begin{document}
\begin{center}
\vspace{1.cm}
{\bf \large Trilinear Gauge Interactions in Extensions of the Standard Model  and Unitarity\footnote{Presented by Roberta Armillis at IFAE 2009, Bari, 15-17 April 2009, Italy}\\ }
\vspace{1.5cm}
{\bf Roberta Armillis, Claudio Corian\`{o}, Luigi Delle Rose  }

\vspace{1cm}

{\it $^a$Dipartimento di Fisica, Universit\`{a} del Salento \\
and  INFN Sezione di Lecce,  Via Arnesano 73100 Lecce, Italy}\\
\vspace{.5cm}
\begin{abstract}
We summarize recent work on the characterization of anomaly poles in connection with the field-theory interpretation of the Green-Schwarz mechanism of anomaly cancellation and on their effective field theories, stressing on the properties of the anomaly vertex in two representations, the Rosenberg and the Longitudinal/Transverse. The presence of polar amplitudes in these theories causes a violation of unitarity at high energy which is cured by the 
exchange of the axion. We comment on the possible physical implications of this mechanism.

\end{abstract}
\end{center}
\newpage

\section{Introduction}
Abelian extensions of the Standard Model represent an economical but yet profound modification of the gauge structure of the electroweak sector, which can be tested at the LHC. Since $U(1)$ interactions abound in effective theories derived from string theory or from Grand Unified Theories (GUT's), establishing the origin of these extensions, if found at the new collider, would be of paramount importance. Several compactifications of string theory predict the existence of anomalous $U(1)$ symmetries and the mechanism of anomaly cancellation in these effective models requires an axion. Understanding at a more phenomenological level the meaning of this cancellation and its implications both in collider experiments and in a cosmological context is rather challenging, since several aspects of this construction remain unclear. We present simple examples to show how challenging this cancellation -from a phenomenological perspective- can be and, if truly realized in nature, how it would provide further insights on the path towards unification. We outline the basic features of our investigation. More details can be found in \cite{Armillis:2009sm}.

\section{Anomalies and their cancellation} 
\begin{figure}[t]
\begin{center}
\includegraphics[scale=0.7]{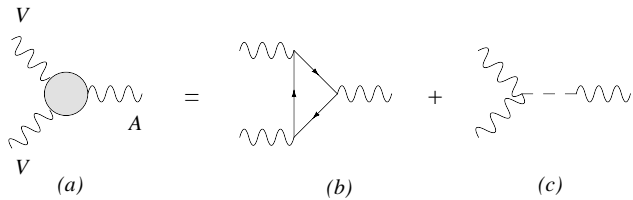}
\caption{\small The diagrammatic form of the GS vertex in the AVV case (a), composed of an AVV triangle (b) and of a single counterterm of polar form with the exchange of the axion (c).}
\label{GS_AVV}
\end{center}
\end{figure}

In an anomaly-free theory, trilinear gauge interactions mediated by anomaly diagrams are identically vanishing. 
In fact, one of the most important aspect of the charges of the Standard Model is that they satisfy cubic relations which strongly constrain the fermion spectrum, which is chiral. We can think of the anomaly diagram as composed of two contributions, the longitudinal $(\Delta_L)$ and the transverse part $(\Delta_T)$. Anomaly cancellation by charge assignment sets to zero both parts of the anomaly vertex. An alternative route, which involves the axion, is to cancel the same dangerous diagrams by an extra contribution. This extra contribution, in an ordinary field theory formulation, consists of a non-local exchange, with the (massless) axion coupled derivatively to the anomalous gauge boson (here denoted by $B$).  Misleadingly, the cancellation is often described as being identical, due to the combination of the two diagrams (b) and (c), shown in Fig.~\ref{GS_AVV}, which would sum up to zero. This is not true, of course, since the exchange of the axion in Fig.~\ref{GS_AVV}c removes only the longitudinal part of the anomaly diagram $\Delta_L$, leaving its transverse component free. 
The vertex represented in Fig.~\ref{GS_AVV}b is indeed
\begin{equation}
 \mathcal \, \Delta ^{\lambda\mu\nu}= \frac{1}{8\pi^2} \left [  \mathcal \, \Delta^{L\, \lambda\mu\nu} -  \mathcal \, \Delta^{T\, \lambda\mu\nu} \right],
\label{long}
\end{equation}
where the longitudinal component
\begin{equation}
 \mathcal \, \Delta^{L\, \lambda\mu\nu}= -\frac{4 i}{k^2}  \, k^\lambda \varepsilon^{\mu \nu \alpha\beta} k_{1 \alpha} 
 k_{2\beta}
\end{equation}
describes the anomaly pole. In these notations $k$ is the incoming momentum of the axial-vector line, while $k_1$ and $k_2$ denote the outcoming momenta of the two vector lines.
Diagram (c) is the pole subtraction mechanism, usually quoted in string theory as ``the Green Schwarz mechanism" (GS). 
Truly, this pole subtraction has appeared in field theory in several cases before and string theory offers just one justification of this cancellation, using the string spectrum. Anomaly diagrams are affected by anomaly poles, but only in some kinematical configuration. We are going to describe briefly this point, leaving the rest of the details to our works on the subject \cite{Armillis:2009sm, Armillis:2008bg}.

 \section{The pole device} 
 We may wonder whether the massless exchange (which is a pole) is an {\em ad hoc} subtraction or there is more to it. For this we need to get back to the anomaly vertex and to its popular description as given by Rosenberg in '63 and found in all the field theory textbooks. A dispersive analysis of this diagram shows that the anomaly diagram is identical to its pole counterterm (i.e. the diagram with the axion) only in a special situation, that is when the two vector lines are on shell. This special kinematic situation (we call it the ``collinear fermion/antifermion limit") is the only one in which the cancellation of the anomaly diagram and of its counterterm is identical. In the opposite case (that we call ``the non-collinear limit"), when the vector lines have nonzero virtualities, the counterterm is not part of the vertex and its introduction may look artificial.
\begin{figure}[t]
\begin{center}
\includegraphics[width=9cm]{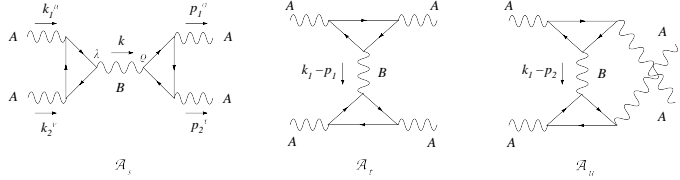}
\caption{\small The scattering process  $AA\rightarrow AA$ via a BIM amplitude in the $s, t, u$ channel mediated by BIM amplitudes}
\label{AAAA}
\end{center}
\end{figure}

\subsection{The infrared pole}
We can summarize the previous observation by saying that the anomaly diagram is pole-dominated only in certain configurations, and if we are away from those, any subtraction of a pole leaves a pole coupled in the infrared. In fact, if we define the effective ``anomaly free" vertex, in field theory, as the sum of the two diagrams 
(see Fig.~\ref{GS_AVV}a) 
and take it as a replacement for the standard triangle diagram, we find that this re-defined vertex, once we are in the collinear limit, is vanishing, and manifests otherwise a pole in the non-collinear case (this is the opposite of what the anomaly diagram provides for us before the subtraction).  An alternative analysis carried out using instead of Rosenberg's form of the anomaly diagram a formulation due to Knecht et al. \cite{Knecht:2003xy} obtained as a solution of the Ward identities of the anomaly vertex, allows to isolate a pole from the same vertex both in the collinear and in the non-collinear case, via a $\Delta_L/\Delta_T$ decomposition at the cost of introducing some extra singularities. This representation can be mapped to Rosenberg's  and a careful analysis shows that both, in the non-collinear limit, are free from anomaly poles \cite{Armillis:2009sm}. However, the L/T representation, which formally isolates a pole (which is contained in $\Delta_L$) for {\em any} momenta of the vertex, is very useful for the analysis in the ultraviolet of a class of dangerous amplitudes that violate unitarity. In simple words: if we don't erase the anomaly vertex, then we need to worry about the exchange of longitudinal components which are associated with their anomaly poles.  
\section{Anomaly poles in different parameterizations and BIM amplitudes}
 The longitudinal components appear in a class of S-matrix elements (named Bouchiat-Iliopoulos-Meyer,
or BIM amplitudes in \cite{Coriano:2008pg}) which break unitarity at high energy. These are compatible with unitarity only by the subtraction of their anomaly poles. They are obtained by sewing together two anomaly vertices in the $s, t$ and $u$ channels (see Fig.~\ref{AAAA}). A computation of these contributions in the simplest case (for on-shell vector lines) gives, in the center of mass frame, for the corresponding squared amplitude  \cite{Armillis:2009sm}
\begin{equation}
| \overline \mathcal  M |^2 _{AA \rightarrow AA} (s, \theta) =
\frac{|a_n|^4}{64 } \frac{s^2}{M_B^4} (\cos^2 \theta + 3)
\label{mathcalM}
\end{equation}
which breaks unitarity at high energy. 
If we use the L/T formulation and interpret the GS mechanism as a subtraction of the polar components  (subtraction which is present both in the collinear and non collinear limits), these amplitudes do not violate unitarity. It is therefore tempting to interpret the pole present in the L/T 
formulation (which is indeed present for any kinematics of the re-defined vertex) as also describing an ultraviolet effect, due to its appearance in a longitudinal exchange at high energy. It can be shown \cite{Armillis:2009sm} that even in the scattering of massive gauge boson, similar violations of unitarity are encountered. This is somehow unobvious, since there are no pole contributions for {\em massive} gauge lines in each of the two anomaly diagrams of a BIM amplitude. This is a compelling argument to refrain from looking at the polar contributions just as to an infrared effect. We find this not too surprising, though, since there is no renormalization scale dependence of an anomaly diagram, and the separation between UV and IR regions, in the absence of other scales, induced by massive fermions which decouple and leave a light fermion spectrum at low energy, hard to realize.

\section{Conclusions} 
 The pole subtraction is nonlocal (see the discussion in \cite{Coriano:2008pg}), but can be cast into a local form by introducing auxiliary fields. Similar auxiliary fields have been found recently in the description of the trace anomaly in gravity and claimed to be fundamental \cite{Giannotti:2008cv}. These fields are, however, puzzling, since one of them is a ghost, having a negative kinetic energy. 
 We don't think that there are unique conclusions regarding the physical interpretation of these poles, as far as anomalous gauge theories are concerned. If the counterterms are engineered to cancel them -in such a way to remove the non unitary growth of BIM amplitudes in these theories- then the re-defined vertex of Fig.~\ref{GS_AVV}a has an infrared pole coupled in the non-collinear limit and should somehow disappear from the (massless) spectrum. If we assume that these gauge theories 
 have a non-perturbative phase, one could use the analogy to chiral theories to claim that the pole can be made massive in this phase, as in the pion case.  A second possibility is that a consistent coupling of these models to gravity may cause a cancellation of these poles and of the corresponding BIM amplitudes of other sectors.
 The phenomenological implications of these studies are interesting, and may bring us to a more complete understanding of the role of anomalous $U(1)$ symmetries and of light pseudoscalars and/or moduli from string theory in the early universe. Supersymmetric extensions of these models have also been formulated recently \cite{Coriano:2008xa}. Open issues are the investigation of the gauging of supersymmetry to obtain special forms of supergravities containing axionic symmetries which have been studied recently \cite{DeRydt:2007vg}. Unitarity sets out strong consistency conditions 
which can be inferred in a bottom-up approach and are very useful to bring these theories closer to experiments.

\centerline {\bf Acknowledgments}
We thank Marco Guzzi for collaborating to this analysis and Antonio Mariano for discussions.
This work was supported (in part) by the European Union through the Marie Curie Research and Training Network ``Universenet'' (MRTN-CT-2006-035863).

\end{document}